\documentclass[a4paper,12pt]{report}
\usepackage{amsmath}
\makeindex
\begin{document}
\parindent=1.05cm
\setlength{\baselineskip}{14truept} \setcounter{page}{1}
\makeatletter
\begin{center}
\Large{\bf The superkinetic and interacting terms of
Chiralsuperfields}
\end{center}
\begin{center}
P. T. Tuyen \footnote {Email: tuyenpt@coltech.vnu.vn}; H. V. Vinh \footnote{Email: vanvinhoang@gmail.com}\\
\textit{Department of Theoretical Physics, Faculty of Physics, Hanoi University of Science, 334
Nguyen Trai, Thanh xuan, Hanoi, Vietnam}
\end{center}
\begin{center}
\hspace*{20pt} \textit{In this paper, we achieve some interesting
results in the way to make sense how superparticles interact
together and to ordinary particles by means of putting aside the
dimensional constraints. This is the first step in the process of
constructing an effective model taking binding possibilities of
superparticles into account.
}

\end{center}

{\bf I. In brief}
\vskip0.2cm
 In hadron models, we see that quarks
and antiquarks combine together and form bound-states as baryons and
mesons. Logically, it is possible that their superpartners combine
in somehow and form something by the name of superbarryons and
supermesons. As we know, squarks and anti-squarks are scalars.
Hence, superbarryons and supermesons are scalars too. For the time
being, this topic has not been mentioned, because general
supersymmetric gauge field theories have not been completely
constructed yet.  Because of complicated calculations, physicists
usually put constraints into superfields in order to get some
interacting initial terms of interacting Lagrangian when
constructing supersymmetric gauge field theories. Unfortunately,
several interacting possibilities between particles and sparticles
are missed due to these constraints. So that, we can not predict
some new particles which may pay important roles in
effective models.
 In the framework of this paper, we put some constraints on
supersymmetric Lagrangian aside with a hope of finding more
interacting possibilities between particles and sparticles. From
these results, we expect to contribute to construct a better
phenomenological theory for mesons and nucleons.

\vskip0.2cm
{\bf II. Superkinetic of a chiral superfield and interacting terms}
\vskip0.2cm
{\bf 2.1 Superkinetic terms}
\vskip0.2cm

In Supersymmetric field
theories,supersymmetric Lagrangians are usually constructed by using
supersymmetric utilities. It is seemingly undoubted. By more
strictly considering, however, we realize that these Lagrangians are
constructed from the parts whose have potential characteristic. The
parts consist of two members f and K. Here, f is called
superpotential, K is Kahler potential. Moreover, whether the super
derivatives are defined for supersymmetric transformations? These
are the motivation for us to live no stone unturned to obtain new
contributions by which are
produced using these superderivatives

\vskip0.2cm
{\bf 2.1.1 Fundamental blocks }
\vskip0.2cm

\begin{equation}
 D_\alpha  \left( {\Phi \left( {x,\theta } \right)} \right) =  -
\left( {\frac{\partial } {{\partial \bar \theta _\alpha  }} + \left(
{\gamma ^\mu  \theta } \right)_\alpha  \partial _\mu  }
\right)\left( {\Phi \left( {x,\theta } \right)} \right)
\end{equation}
\[
 = \sqrt 2 \,\left( {\psi _L } \right)_\alpha   - \left( {\gamma ^\mu  \theta } \right)_\alpha  \partial _\mu  \varphi  - \theta _\alpha  H - \left( {\gamma _5 \theta } \right)_\alpha  H - \left( {\gamma _5 \gamma ^\mu  \theta } \right)_\alpha  \partial _\mu  \varphi
 +\]
 \[+ \sqrt 2 \left( {\gamma ^\mu  \theta } \right)_\alpha  \left( {\bar \theta \partial _\mu  \psi _L } \right) + \sqrt 2 \left( {\gamma _5 \theta } \right)_\alpha  \left( {\bar \theta \gamma ^\mu  \partial _\mu  \psi _L } \right) + \frac{1}
{{\sqrt 2 }}\left( {\bar \theta \gamma _5 \theta } \right)\left(
{\gamma ^\mu  \partial _\mu  \psi _L } \right)_\alpha   -\]
\[ - \frac{1}
{2}\left( {\gamma ^\mu  \theta } \right)_\alpha  \left( {\bar \theta
\theta } \right)\partial _\mu  H - \frac{1} {2}\left( {\gamma ^\mu
\theta } \right)_\alpha  \left( {\bar \theta \gamma _5 \theta }
\right)\partial _\mu  H - \frac{1} {2}\left( {\gamma ^\mu  \theta }
\right)_\alpha  \left( {\bar \theta \gamma _5 \gamma ^\nu  \theta }
\right)\partial _\mu  \partial _\nu  \varphi  +\]
\[ + \frac{1}
{2}\left( {\gamma _5 \theta } \right)_\alpha  \left( {\bar \theta
\gamma _5 \theta } \right)\partial ^\mu  \partial _\mu  \varphi  +
\frac{{\sqrt 2 }} {2}\left( {\gamma ^\mu  \theta } \right)_\alpha
\left( {\bar \theta \gamma _5 \theta } \right)\left( {\bar \theta
\gamma ^\nu  \partial _\mu  \partial _\nu  \psi _L } \right)\]
\begin{equation}
\bar DD\left( {\Phi \left( {x,\theta } \right)} \right) =  - 4H +
2\sqrt 2 \bar \theta \gamma ^\mu  \partial _\mu  \psi _L  + 2\sqrt 2
\bar \theta \gamma ^\mu  \Upsilon _5 \partial _\mu  \psi _L  +
\end{equation}
\[
 + 2\left( {\bar \theta \gamma _5 \gamma ^\mu  \theta } \right)\partial _\mu  H - 2\left( {\bar \theta \theta } \right)\partial ^\mu  \partial _\mu  \varphi  + 2\left( {\bar \theta \gamma _5 \theta } \right)\partial ^\mu  \partial _\mu  \varphi  + \sqrt 2 \left( {\bar \theta \theta } \right)\left( {\bar \theta \partial ^\mu  \partial _\mu  \psi _L } \right) -
\]
\[
 - \sqrt 2 \left( {\bar \theta \gamma _5 \theta } \right)\left( {\bar \theta \partial ^\mu  \partial _\mu  \psi _L } \right) + \frac{1}
{2}\left( {\bar \theta \gamma _5 \theta } \right)^2 \partial ^\mu
\partial _\mu  H - \frac{1} {2}\left( {\bar \theta \theta }
\right)\left( {\bar \theta \gamma _5 \theta } \right)\partial ^\mu
\partial _\mu  H
\]
\begin{equation}
D_\alpha  \left( {\Phi ^* \left( {x,\theta } \right)} \right) =
\sqrt 2 \left( {\psi _R } \right)_\alpha   - \left( {\gamma ^\mu
\theta } \right)_\alpha  \partial _\mu  \varphi ^*  - \theta _\alpha
H^*  + \left( {\gamma _5 \theta } \right)_\alpha  H^*  +
\end{equation}
\[
 + \left( {\gamma _5 \gamma ^\mu  \theta } \right)_\alpha  \partial _\mu  \varphi ^*  + \sqrt 2 \left( {\gamma ^\mu  \theta } \right)_\alpha  \left( {\bar \theta \partial _\mu  \psi _R } \right) - \sqrt 2 \left( {\gamma _5 \theta } \right)_\alpha  \left( {\bar \theta \gamma ^\mu  \partial _\mu  \psi _R } \right) -
\]
\[
 - \frac{1}
{{\sqrt 2 }}\left( {\bar \theta \gamma _5 \theta } \right)\left(
{\gamma ^\mu  \partial _\mu  \psi _R } \right)_\alpha   - \frac{1}
{2}\left( {\gamma ^\mu  \theta } \right)_\alpha  \left( {\bar \theta
\theta } \right)\partial _\mu  H^*  + \frac{1} {2}\left( {\gamma
^\mu  \theta } \right)_\alpha  \left( {\bar \theta \gamma _5 \theta
} \right)\partial _\mu  H^*  +
\]
\[
 + \frac{1}
{2}\left( {\gamma ^\mu  \theta } \right)_\alpha  \left( {\bar \theta
\gamma _5 \gamma ^\nu  \theta } \right)\partial _\mu  \partial _\nu
\varphi ^*  + \frac{1} {2}\left( {\gamma _5 \theta } \right)_\alpha
\left( {\bar \theta \gamma _5 \theta } \right)\partial ^\mu
\partial _\mu  \varphi ^*  - \frac{1} {{\sqrt 2 }}\left( {\gamma
^\mu  \theta } \right)_\alpha  \left( {\bar \theta \gamma _5 \theta
} \right)\left( {\bar \theta \gamma ^\nu  \partial _\mu  \partial
_\nu  \psi _R } \right)
\]
\begin{equation}
\bar DD\left( {\Phi ^* \left( {x,\theta } \right)} \right) =  - 4H^*
+ 2\sqrt 2 \bar \theta \gamma ^\mu  \partial _\mu  \psi _R  + 2\sqrt
2 \bar \theta \gamma ^\mu  \gamma _5 \partial _\mu  \psi _R  -
\end{equation}
\[
 - 2\left( {\bar \theta \gamma _5 \gamma ^\mu  \theta } \right)\partial _\mu  H^*  - 2\left( {\bar \theta \theta } \right)\partial ^\mu  \partial _\mu  \varphi ^*  - 2\left( {\bar \theta \gamma _5 \theta } \right)\partial ^\mu  \partial _\mu  \varphi ^*  + \sqrt 2 \left( {\bar \theta \theta } \right)\left( {\bar \theta \partial ^\mu  \partial _\mu  \psi _R } \right) -
\]
\[
 - \sqrt 2 \left( {\bar \theta \gamma _5 \theta } \right)\left( {\bar \theta \partial ^\mu  \partial _\mu  \psi _R } \right) + \frac{1}
{2}\left( {\bar \theta \gamma _5 \theta } \right)^2 \partial ^\mu
\partial _\mu  H^*  - \frac{1} {2}\left( {\bar \theta \theta }
\right)\left( {\bar \theta \gamma _5 \theta } \right)\partial ^\mu
\partial _\mu  H^*
\]
\begin{equation}
\bar D_\alpha  \left( {\Phi \left( {x,\theta } \right)} \right) =
\sqrt 2 \left( {\bar \psi _L } \right)_\alpha   + \left( {\bar
\theta \gamma ^\mu  } \right)_\alpha  \partial _\mu  \varphi  - \bar
\theta _\alpha  H - \left( {\bar \theta \gamma _5 } \right)_\alpha
H -
\end{equation}
\[
 - \left( {\bar \theta \gamma _5 \gamma ^\mu  } \right)_\alpha  \partial _\mu  \varphi  - \sqrt 2 \left( {\bar \theta \gamma ^\mu  } \right)_\alpha  \left( {\partial _\mu  \bar \psi _L \theta } \right) - \sqrt 2 \left( {\bar \theta \gamma _5 } \right)_\alpha  \left( {\partial _\mu  \bar \psi _L \gamma ^\mu  \theta } \right) -
\]
\[
 - \frac{1}
{{\sqrt 2 }}\bar \theta \gamma _5 \theta \left( {\partial _\mu  \bar
\psi _L \gamma ^\mu  } \right)_\alpha   + \frac{1} {2}\left( {\bar
\theta \gamma ^\mu  } \right)_\alpha  \left( {\bar \theta \theta }
\right)\partial _\mu  H + \frac{1} {2}\left( {\bar \theta \gamma
^\mu  } \right)_\alpha  \left( {\bar \theta \gamma _5 \theta }
\right)\partial _\mu  H +
\]
\[
 + \frac{1}
{2}\left( {\bar \theta \gamma ^\mu  } \right)_\alpha  \left( {\bar
\theta \gamma _5 \gamma ^\nu  \theta } \right)\partial _\mu
\partial _\nu  \varphi  + \frac{1} {2}\left( {\bar \theta \gamma _5
} \right)_\alpha  \left( {\bar \theta \gamma _5 \theta }
\right)\partial ^\mu  \partial _\mu  \varphi  + \frac{{\sqrt 2 }}
{2}\left( {\bar \theta \gamma ^\mu  } \right)_\alpha  \left( {\bar
\theta \gamma _5 \theta } \right)\left( {\partial _\mu  \partial
_\nu  \bar \psi _L \gamma ^\nu  \theta } \right)
\]
\begin{equation}
\bar D_\alpha  \left( {\Phi ^* \left( {x,\theta } \right)} \right) =
\sqrt 2 \left( {\bar \psi _R } \right)_\alpha   + \left( {\bar
\theta \gamma ^\mu  } \right)_\alpha  \partial _\mu  \varphi ^*  -
\bar \theta _\alpha  H^*  + \left( {\bar \theta \gamma _5 }
\right)_\alpha  H^*  +
\end{equation}
\[
 + \left( {\bar \theta \gamma _5 \gamma ^\mu  } \right)_\alpha  \partial _\mu  \varphi ^*  - \sqrt 2 \left( {\bar \theta \gamma ^\mu  } \right)_\alpha  \left( {\partial _\mu  \bar \psi _R \theta } \right) - \sqrt 2 \left( {\bar \theta \gamma _5 } \right)_\alpha  \left( {\partial _\mu  \bar \psi _R \gamma ^\mu  \theta } \right) +
\]
\[
 + \frac{1}
{{\sqrt 2 }}\left( {\bar \theta \gamma _5 \theta } \right)\left(
{\partial _\mu  \bar \psi _R \gamma ^\mu  } \right)_\alpha   +
\frac{1} {2}\left( {\bar \theta \gamma ^\mu  } \right)_\alpha
\left( {\bar \theta \theta } \right)\partial _\mu  H^*  - \frac{1}
{2}\left( {\bar \theta \gamma ^\mu  } \right)_\alpha  \left( {\bar
\theta \gamma _5 \theta } \right)\partial _\mu  H^*  -
\]
\[
 - \frac{1}
{2}\left( {\bar \theta \gamma ^\mu  } \right)_\alpha  \left( {\bar
\theta \gamma _5 \gamma ^\nu  \theta } \right)\partial _\mu
\partial _\nu  \varphi ^*  + \frac{1} {2}\left( {\bar \theta \gamma
_5 } \right)_\alpha  \left( {\bar \theta \gamma _5 \theta }
\right)\partial ^\mu  \partial _\mu  \varphi ^*  + \frac{1} {{\sqrt
2 }}\left( {\bar \theta \gamma ^\mu  } \right)_\alpha  \left( {\bar
\theta \gamma _5 \theta } \right)\left( {\partial _\mu  \partial
_\nu  \bar \psi _R \gamma ^\nu  \theta } \right)
\]

\vskip0.2cm
{\bf 2.1.2 Combinations}
\vskip0.2cm

These are some basic objects.We will imply abilities to combine them
in order to contribute new parts into Lagrangian. The general action
is added by term as:
\begin{equation}
I_H  = \frac{1} {2}\int {d^4 x\left( {H_D^2 } \right)}  + \frac{1}
{8}\int {d^4 x\left( {H_D^4 } \right)}
\end{equation}
Where:
\begin{equation}
H_D^2  = \left[ {\bar D\left( {\Phi ^* \left( {x,\theta } \right)}
\right)D\left( {\Phi \left( {x,\theta } \right)} \right) + \bar
D\left( {\Phi ^* \left( {x,\theta } \right)} \right)\gamma _5
D\left( {\Phi \left( {x,\theta } \right)} \right)} \right]_{\bar
\theta \bar \theta \theta \theta }
\end{equation}
and
\begin{equation}
H_D^4  = \left[ {\bar DD\left( {\Phi ^* \left( {x,\theta } \right)}
\right)\bar DD\left( {\Phi \left( {x,\theta } \right)} \right)}
\right]_{\bar \theta \bar \theta \theta \theta }
\end{equation}
After meticulous calculations, we achieve some exciting results:
\begin{equation}
H_D^2  = 0
\end{equation}
\begin{equation}
H_D^4  = 2\partial ^\mu  \partial _\mu  \left( {H^* H} \right) -
8\partial ^\mu  H^* \partial _\mu  H + 8\partial ^\mu  \partial _\mu
\varphi ^* \partial ^\mu  \partial _\mu  \varphi
\end{equation}
In this expression, the first term is a total derivative. Hence, it
vanish when is put in the action. While, the third term, the
unexpected term, will be canceled by on-shell condition and massless
scenario. That means:
\begin{equation}
\partial ^\mu  \partial _\mu  \varphi ^*  = 0;\,\,\partial ^\mu  \partial _\mu  \varphi  = 0
\end{equation}
\begin{equation}
H_D^4  = \partial ^\mu  H^* \partial _\mu  H
\end{equation}
Hence, we can write new Lagrangian density:
\begin{equation}
L_H  = \partial ^\mu  H^* \partial _\mu  H
\end{equation}
\vskip0.2cm
{\bf 2.2 Interacting terms }
\vskip0.2cm
we expect to determine the interacting forms between sparticles with
ordinary particles. Let consider N superfields $ \Phi _n(x,\theta)$
and N complex conjugate superfields $ \Phi _n^* (x,\theta )$. They
are expressed as:
\begin{equation}
\Phi _n \left( {x,\theta } \right) = \varphi _n \left( {x_ +  }
\right) - \sqrt 2 \theta _L^T \varepsilon \psi _{Ln} \left( {x_ +  }
\right) + H_n \left( {x_ +  } \right)\left( {\theta _L^T \varepsilon
\theta _L } \right)
\end{equation}
\begin{equation}
\tilde \Phi _n \left( {x,\theta } \right) = \tilde \varphi _n \left(
{x_ -  } \right) - \sqrt 2 \theta _R^T \varepsilon \psi _{Rn} \left(
{x_ -  } \right) + \tilde H_n \left( {x_ -  } \right)\left( {\theta
_L^T \varepsilon \theta _L } \right)
\end{equation}
While, ${x_ +  }$ and ${x_ -  } $ are
\begin{equation}
x_ \pm ^\mu   = x^\mu   \pm \theta _R^T \varepsilon \gamma ^\mu
\theta _L
\end{equation}
The interacting function is performed as formula:
\begin{equation}
f\left( \Phi  \right) = \sum\limits_{i = 1}^N {f_i \left( {\Phi _i }
\right)}  + t\sum\limits_{i \ne j = 1}^N {\Phi _i \Phi _j }  +
d\sum\limits_{i \ne j \ne k = 1}^N {\Phi _i \Phi _j \Phi _k }  + h.c
\end{equation}
Correspondingly, the interacting action is
\begin{equation}
\left. {I_{\operatorname{int} }  = \int {dx^4 \left\{
{\sum\limits_{i = 1}^N {\left[ {f_i \left( {\Phi _i } \right)}
\right]_F } } \right.}  + t\sum\limits_{i \ne j = 1}^N {\left[ {\Phi
_i \Phi _j } \right]_F }  + d\sum\limits_{i \ne j \ne k = 1}^N
{\left[ {\Phi _i \Phi _j \Phi _k } \right]_F }  + h.c} \right\}
\end{equation}
In detail
\begin{equation}
\left[ {f_i \left( {\Phi _i } \right)} \right]_F  = H_i
\frac{{\partial f_i }} {{\partial \varphi _i }} - \frac{1}
{2}\frac{{\partial ^2 f_i }} {{\partial \varphi _i^2 }}\bar \psi
_{Li} \psi _{Li} {\text{ }}
\end{equation}
\begin{equation}
\left[ {\Phi _i \Phi _j } \right]_F {\text{ =  }}\varphi _{\text{i}}
{\text{H}}_{\text{j}} {\text{ + }}\varphi _{\text{j}}
{\text{H}}_{\text{i}} {\text{ - }}\bar \psi _{{\text{Li}}} \psi
_{{\text{Lj}}} {\text{ }}
\end{equation}
\begin{equation}
\left[ {\Phi _i \Phi _j \Phi _k } \right]_F  = \varphi _i \varphi _j
H_k  + \varphi _j \varphi _k H_i  + \varphi _k \varphi _i H_j  -
\hfill \\
\end{equation}
\[-\bar \psi _{Li} \psi _{Lj} \varphi _k  - \bar \psi _{Lj} \psi _{Lk} \varphi _i  - \bar \psi _{Li} \psi _{Lk} \varphi _j  \hfill \\
\]
Remember that$\varphi _i ,\psi _{Li} ,\bar \psi _{Li} ,H_i $, depend
on $x_ +   = x^\mu   + \theta _R^T \varepsilon \gamma ^\mu  \theta
_L $ . However, it is possible to achieve the dependence of them on
Minkovskian variables through a transformation in the action. Hence,
we have:
\begin{equation}
I_{\operatorname{int} }  = I_i  + I_{ijk}
\end{equation}
Where
\begin{equation}
I_i  = \int {dx^4 } \left\{ {\sum\limits_{i = 1}^N {\left[ {H_i
\frac{{\partial f_i }} {{\partial \varphi _i }} - \frac{1}
{2}\frac{{\partial ^2 f_i }} {{\partial \varphi _i^2 }}\bar \psi
_{Li} \psi _{Li} } \right]} } \right\}
\end{equation}
\begin{equation}
I_{ijk}  = \int {dx^4 } \left\{ {\sum\limits_{i \ne j \ne k = 1}^N
{\left[ {t\left( {\varphi _{\text{i}} {\text{H}}_{\text{j}} {\text{
+ }}\varphi _{\text{j}} {\text{H}}_{\text{i}} {\text{ - }}\bar \psi
_{{\text{Li}}} \psi _{{\text{Lj}}} } \right) + d\left( {\varphi
_{[i} \varphi _j H_{k]}  - \bar \psi _{L[i} \psi _{Lj} \varphi _{k]}
} \right)} \right]}  + h.c} \right\}
\end{equation}
We denoted that
\begin{equation}
\varphi _{[i} \varphi _j H_{k]}  = \varphi _i \varphi _j H_k  +
\varphi _j \varphi _k H_i  + \varphi _k \varphi _i H_j
\end{equation}
\[
\bar \psi _{L[i} \psi _{Lj} \varphi _{k]}  = \bar \psi _{Li} \psi
_{Lj} \varphi _k  + \bar \psi _{Lj} \psi _{Lk} \varphi _i  + \bar
\psi _{Li} \psi _{Lk} \varphi _j
\]

\vskip0.2cm
{\bf III. The three superfield systerm }
\vskip0.2cm
We take into account three chiral superfields: quark up superfield
U, quark down superfield D, quark strange superfield S. They are
expressed as:
\begin{equation}
\Phi _i \left( {x,\theta } \right) = \varphi _i \left( {x_ +  }
\right) - \sqrt 2 \theta _L^T \varepsilon \psi _{Li} \left( {x_ +  }
\right) + H_i \left( {x_ +  } \right)\theta _L^T \varepsilon \theta
_L
\end{equation}
Or:
\begin{equation}
\Phi _i \left( {x,\theta } \right) = \varphi _i \left( x \right) -
\,\sqrt 2 \left( {\bar \theta \psi _{Li} \left( x \right)} \right) +
\left( {\bar \theta \left( {\frac{{1 + \gamma _5 }} {2}}
\right)\theta } \right)H_i \left( x \right) +
\end{equation}
\[
 + \frac{1}
{2}\left( {\bar \theta \gamma _5 \gamma ^\mu  \theta }
\right)\partial _\mu  \varphi _i \left( x \right) - \frac{1} {{\sqrt
2 }}\left( {\bar \theta \gamma _5 \theta } \right)\left( {\bar
\theta \not \partial \psi _{Li} \left( x \right)} \right) - \frac{1}
{8}\left( {\bar \theta \gamma _5 \theta } \right)^2 \partial ^\mu
\partial _\mu  \varphi _i \left( x \right)
\]
Here, i=1, 2, 3. $\Phi _{1,2,3}  = U,D,S$ are chiral superfields;
$\psi _{L1,2,3}  = u,d,s$ are quark up, down and strange;$\varphi
_{1,2,3}  = \tilde u,\tilde d,\tilde s$  are superpartners of
quarks; $H_i$ are auxiliary fields of superfields $U,D,S$. The
supersymmetric action:
\begin{equation}
I = I_K  + I_H  + I_{\operatorname{int} }
\end{equation}
Where
\begin{equation}
I_K  = \int {dx^4 \left[ K \right]_D }
\end{equation}
\begin{equation}
I_H  = \frac{{h_1 }} {2}\int {d^4 x\left( {H_D^2 } \right)}  +
\frac{{h_2 }} {8}\int {d^4 x\left( {H_D^4 } \right)}
\end{equation}
\begin{equation}
I_{\operatorname{int} }  = \int {dx^4 } \left\{ {\sum\limits_{i =
1}^3 {\left[ {f_i \left( {\Phi _i } \right)} \right]_F }  +
\sum\limits_{i \ne j = 1}^3 {\eta _{ij} \left[ {\Phi _i \Phi _j }
\right]_F }  + \sum\limits_{i \ne j \ne k = 1}^3 {\varpi _{ijk}
\left[ {\Phi _i \Phi _j \Phi _k } \right]_F } } \right\} +
\end{equation}
\[
\begin{gathered}
  \,\,\,\,\,\,\,\,\, + \int {dx^4 \left\{ {\sum\limits_{i = 1}^3 {\left[ {f_i \left( {\Phi _i } \right)} \right]_F^* }  + \sum\limits_{i \ne j = 1}^3 {\eta _{ij}^* \left[ {\Phi _i \Phi _j } \right]_F^* }  + \sum\limits_{i \ne j \ne k = 1}^3 {\varpi _{ijk}^* \left[ {\Phi _i \Phi _j \Phi _k } \right]_F^* } } \right\} + }  \hfill \\
  \,\,\,\,\,\,\,\,\, + hc \hfill \\
\end{gathered}
\]
$h_1 ,h_2$ are constants with dimensions of -1 and -2, respectively.
In this case, we put renormalization aside, so that these minus
values are acceptable. $\eta _{ij} ,\varpi _{ijk} $ are interacting
constants.\\
Kahler potential, here, is formed as:
\begin{equation}
K(\Phi ^* ,\Phi ) = \sum\limits_{i,j = 1}^3 {g_{{\text{ij}}} \Phi
_i^* \Phi _j }
\end{equation}
\begin{equation}
\left[ {K(\Phi ^* ,\Phi )} \right]_D  = \sum\limits_{i,j = 1}^3
{g_{{\text{ij}}} \left[ { - \partial ^\mu  \varphi _i^* \partial
_\mu  \varphi _j  + H_i^* H_j  + \frac{1} {2}\left( {\partial _\mu
\bar \psi _i \gamma ^\mu  \psi _j } \right) - \frac{1} {2}\left(
{\bar \psi _i \gamma ^\mu  \partial _\mu  \psi _j } \right)}
\right]}
\end{equation}
In order that the kinetic terms of scalar fields and spinor fields
coincide with quantum commutative and anti-commutative rules, then
must be positive. Hence, we choice $g_{{\text{ij}}}  = \delta
_{{\text{ij}}}$
\begin{equation}
\left[ {K(\Phi ^* ,\Phi )} \right]_D  = \sum\limits_{i = 1}^3
{\left[ { - \partial ^\mu  \varphi _i^* \partial _\mu  \varphi _i  +
H_i^* H_i  + \frac{1} {2}\left( {\partial _\mu  \bar \psi _i \gamma
^\mu  \psi _i } \right) - \frac{1} {2}\left( {\bar \psi _i \gamma
^\mu  \partial _\mu  \psi _i } \right)} \right]}
\end{equation}
\begin{equation}
I_K  = \int {dx^4 \left\{ {\sum\limits_{i = 1}^3 {\left[ {\frac{1}
{2}\left( {\partial _\mu  \bar \psi _i \gamma ^\mu  \psi _i }
\right) - \frac{1} {2}\left( {\bar \psi _i \gamma ^\mu  \partial
_\mu  \psi _i } \right) - \partial ^\mu  \varphi _i^* \partial _\mu
\varphi _i  + H_i^* H_i } \right]} } \right\}}
\end{equation}
In $I_{\operatorname{int} }$
\begin{equation}
f_i \left( {\Phi _i } \right) = a_i \Phi _i^3  + b_i \Phi _i^2  +
c_i \Phi _i
\end{equation}
Here, $a_i ,b_i ,c_i $ are constants calculated vie Supersymmetry
Spontaneously Broken
\begin{equation}
I_{\operatorname{int} }  = \int {dx^4 } \left\{ {\sum\limits_{i =
1}^3 {\left[ {a_i H_i  + 2b_i H_i \varphi _i  - b_i \bar \psi _{Li}
\psi _{Li}  + 3c_i H_i \varphi _i^2  - 3c_i \varphi _i \bar \psi
_{Li} \psi _{Li} } \right]} } \right. +
\end{equation}
\[
\begin{gathered}
  \,\,\,\,\,\,\,\,\,\,\,\,\, + \sum\limits_{i = 1}^3 {\left[ {a_i H_i^*  + 2b_i H_i^* \varphi _i^*  - b_i \left( {\bar \psi _{Li} \psi _{Li} } \right)^*  + 3c_i H_i^* \left( {\varphi _i^2 } \right)^*  - 3c_i \varphi _i^* \left( {\bar \psi _{Li} \psi _{Li} } \right)^* } \right]}  +  \hfill \\
  \,\,\,\,\,\,\,\,\,\,\,\,\, + \sum\limits_{i \ne j = 1}^3 {\left[ {\eta _{ij} \left[ {H_i \varphi _j  + H_j \varphi _i  - \bar \psi _{Li} \psi _{Lj} } \right] + \eta _{ij}^* \left[ {H_i^* \varphi _j^*  + H_j^* \varphi _i^*  - \left( {\bar \psi _{Li} \psi _{Lj} } \right)^* } \right]} \right]}  +  \hfill \\
  \,\,\,\,\,\,\,\,\,\,\,\,\,\left. { + \sum\limits_{i \ne j \ne k = 1}^3 {\left[ {\varpi _{ijk}^* \left[ {H_{[i} \varphi _j \varphi _{k]}  - \varphi _{[i} \bar \psi _{Lj} \psi _{Lk]} } \right] + \varpi _{ijk}^* \left[ {\left( {H_{[i} \varphi _j \varphi _{k]} } \right)^*  - \left( {\varphi _{[i} \bar \psi _{Lj} \psi _{Lk]} } \right)^* } \right]} \right]} } \right\} \hfill \\
\end{gathered}
\]
We achieve the Lagrangian:
\begin{equation}
L = \sum\limits_{i = 1}^3 {\left[ {\frac{1} {2}\left( {\partial _\mu
\bar \psi _i \gamma ^\mu  \psi _i } \right) - \frac{1} {2}\left(
{\bar \psi _i \gamma ^\mu  \partial _\mu  \psi _i } \right) -
\partial ^\mu  \varphi _i^* \partial _\mu  \varphi _i  + h\partial
^\mu  H_i^* \partial _\mu  H_i } \right]}  +
\end{equation}
\[
\begin{gathered}
  \,\,\,\,\,\,\,\,\,\,\, + \sum\limits_{i = 1}^3 {\left[ {a_i H_i  + a_i H_i^*  + H_i^* H_i  + 2b_i \left[ {H_i \varphi _i  + H_i^* \varphi _i^*  - \frac{1}
{2}\bar \psi _{Li} \psi _{Li}  - \frac{1}
{2}\left( {\bar \psi _{Li} \psi _{Li} } \right)^* } \right]} \right]}  +  \hfill \\
  \,\,\,\,\,\,\,\,\,\,\,\, + \sum\limits_{i = 1}^3 {3c_i \left[ {H_i \varphi _i^2  - \varphi _i \bar \psi _{Li} \psi _{Li}  + H_i^* \left( {\varphi _i^2 } \right)H^*  - \varphi _i^* \left( {\bar \psi _{Li} \psi _{Li} } \right)^* } \right]}  +  \hfill \\
  \,\,\,\,\,\,\,\,\,\,\,\, + \sum\limits_{i \ne j = 1}^3 {\left[ {\eta _{ij} \left( {H_i \varphi _j  + H_j \varphi _i  - \bar \psi _{Li} \psi _{Lj} } \right) + \eta _{ij}^* \left( {H_i^* \varphi _j^*  + H_j^* \varphi _i^*  - \left( {\bar \psi _{Li} \psi _{Lj} } \right)^* } \right)} \right]}  +  \hfill \\
  \,\,\,\,\,\,\,\,\,\,\,\, + \sum\limits_{i \ne j \ne k = 1}^3 {\left[ {\varpi _{ijk} \left( {H_{[i} \varphi _j \varphi _{k]}  - \varphi _{[i} \bar \psi _{Lj} \psi _{Lk]} } \right) + \varpi _{ijk}^* \left( {\left( {H_{[i} \varphi _j \varphi _{k]} } \right)^*  - \left( {\varphi _{[i} \bar \psi _{Lj} \psi _{Lk]} } \right)^* } \right)} \right]}  \hfill \\
\end{gathered}
\]
Let consider terms at the last line in detail. We expect to get the
reflection of how component fields interact together. We can rewrite
these terms:
\begin{equation}
L_{\operatorname{int} }  = \omega _{uds} H_u \tilde d\tilde s +
\omega _{dus} H_d \tilde u\tilde s + \omega _{sud} H_s \tilde
u\tilde d -
\end{equation}
\[\,\,\,\,\,\,\,\,\, - \omega _{uds} \tilde uds - \omega _{dus} \tilde dus - \omega _{sud} \tilde sud -
cc\]
\vskip0.2cm
{\bf IV. Conclusion }
\vskip0.2cm
By putting the dimensional constraints aside, we achieve some
exciting results, concerning with the existing possibility of $H $,
and how ordinary particles and sparticles and $H $ interact
together. 
\end{document}